\documentclass[sigconf]{acmart}

\AtBeginDocument{%
  }

\setcopyright{acmlicensed}
\copyrightyear{2026}
\acmYear{2026}
\acmDOI{XXXXXXX.XXXXXXX}

\acmBooktitle{SIGMOD Record}
\acmISBN{}

\usepackage{xcolor}
\usepackage{amsmath}
\usepackage{xspace}

\newcommand{\nlx}{\textit{NL-to-X}\xspace}
\newtheorem{definition}{Definition}

\begin{document}
\pagestyle{empty}

\title{Natural Language to What? A Vision for Answer-Object Formation in NL-to-X Querying}

\author{Shengqi Li}
\email{shl142@ucsd.edu}
\affiliation{%
  \institution{University of California San Diego}
  \city{La Jolla}
  \state{California}
  \country{USA}
}

\author{Subhasis Dasgupta}
\email{sudasgupta@ucsd.edu}
\affiliation{%
  \institution{University of California San Diego}
  \city{La Jolla}
  \state{California}
  \country{USA}
}

\author{Ilkay Altintas}
\email{ialtintas@ucsd.edu}
\affiliation{%
  \institution{University of California San Diego}
  \city{La Jolla}
  \state{California}
  \country{USA}
}

\author{Amarnath Gupta}
\orcid{0000-0003-0897-120X}
\email{a1gupta@ucsd.edu}
\affiliation{%
  \institution{University of California San Diego}
  \city{La Jolla}
  \state{California}
  \country{USA}
}

\renewcommand{\shortauthors}{Li et al.}

\begin{abstract}
Natural-language interfaces to data are usually framed as translation into a predetermined backend language such as SQL, GQL, SPARQL, or a retrieval/query interface. That framing remains appropriate when the semantic target is already known, but it is too narrow for heterogeneous environments where users may ask for evidence structures, comparisons, timelines, conflicts, catalog-discovery objects, or cross-source witnesses. This paper proposes the \nlx lens for such settings. The central question is not only how to translate a natural-language query, but whether the target is fixed, partially specified by a known answer-object family, or must itself be constructed. We develop a compact systems vocabulary for this problem based on environment profiles, target contracts, realization plans, preservation traces, and answer objects. We instantiate the framework through a claim-conflict answer-object family, showing how a query over announcements and internal emails requires claims, issue alignment, conflict relations, witnesses, provenance, uncertainty, and review status rather than retrieval alone. The result is a vision for target-aware natural-language data systems and a research agenda around target-contract derivation, governed answer-object libraries, realization planning, preservation-aware execution, and benchmarks for target fidelity.
\end{abstract}



\maketitle

\section{Introduction}
\label{sec:intro}

Natural-language interfaces to data are usually framed as translation problems. A user poses a natural-language query, and the system maps it into a target expression that can be executed by a data system. In the most familiar case, the target expression is SQL and the backend is a relational database. Analogous ``fixed-target'' formulations arise for graph databases, RDF stores, search engines, and other systems where the destination language is known in advance: Cypher or GQL for property graphs, SPARQL for RDF graphs, keyword or structured retrieval syntax for text indexes, or API calls for tool-mediated systems.

This framing has produced substantial progress in semantic parsing, schema grounding, query synthesis, execution-guided repair, and benchmark-driven evaluation. Its strength comes from a simplifying assumption: the semantic target is already determined. The system may still need to resolve ambiguity, identify schema elements, infer joins, ground constants, select functions, or repair malformed expressions. But the kind of target itself is not in question. The task is natural language to SQL, natural language to SPARQL, natural language to a graph query, or natural language to another declared execution formalism.

Increasingly, however, users pose questions whose answers are not naturally identified with the native result type of a single backend system. A query over emails, reports, metadata records, and structured data may require more than a tuple set, graph path, triple pattern, or ranked list of documents. It may require an object that organizes evidence across sources: a conflict between two claims, a timeline of how an issue evolved over emails and office memos, or a witness linking documents, records, entities, and timestamps. In such cases, the first systems question is not only how to translate the query, but what answer target must be formed so that execution has a well-defined object to produce.

This paper proposes the \nlx lens for this broader class of natural-language-initiated querying. We do not replace NL2SQL, NL-to-GQL, NL-to-SPARQL, retrieval, data integration, or heterogeneous query processing. Those remain essential. Rather, \nlx asks when the target of interpretation is fixed, when it is only partially specified by a known answer-object family, and when the target itself must be constructed. In fixed-target regimes, the destination language and answer form are already available. In partially specified regimes, the system knows the kind of answer object it must produce but still has to instantiate it. In target-construction regimes, even the relevant answer-object family, evidence structure, or catalog vocabulary is not yet fixed by the environment.

The central claim is that target formation is a missing abstraction problem in natural-language data systems. In heterogeneous settings, a system needs more than a generated query or a sequence of tool calls. It needs an inspectable representation of what the answer is supposed to be. We model this layer through five artifacts: an environment profile $\Gamma(E)$ describing available sources and capabilities; a target contract $\tau$ specifying the answer object and its obligations; a realization plan $\pi$ describing how the system attempts to construct the answer; a preservation trace $\rho$ recording what was preserved, approximated, weakened, or left for review; and the returned answer object $a$. Together, these artifacts make it possible to ask not only whether an answer is plausible, but whether it satisfies or is explicitly trace-qualified against the target contract.

This view changes what must be evaluated. In fixed-target systems, execution accuracy, denotational equivalence, and query exact match remain appropriate. In target-construction settings, however, an answer may fail even if it is fluent, relevant, or backed by retrieved passages. It may fail because the wrong answer-object family was selected, required witnesses were omitted, provenance was dropped, uncertainty was hidden, or an approximation was presented as full satisfaction of the target. The evaluation problem therefore includes target-status diagnosis, target-contract adequacy, answer-object construction, witness coverage, provenance preservation, and no-unreported degradation.

We make these ideas concrete through a claim-conflict answer-object family. A query such as \textit{``Find internal emails that contradict the claims of this announcement''} cannot be adequately represented as retrieval alone. It requires reference claims, candidate claims, issue alignment, a conflict relation, source-span witnesses, provenance, and uncertainty. The same target contract may be realized through retrieval, extraction, graph operations, metadata lookup, model-based contradiction scoring, and human review. The important point is that the target contract states what must be preserved before the system chooses how to execute.

\noindent\textbf{Contributions.} This paper makes four contributions.

\begin{enumerate}
\item It identifies \textit{target formation} as a first-class abstraction problem in natural-language-initiated querying over heterogeneous, document-centric, catalog-mediated, and tool-mediated environments.

\item It introduces a \textit{target-status view} that distinguishes fixed-target, partially specified, and target-construction regimes.

\item It develops a compact systems vocabulary for target-aware querying based on environment profiles, target contracts, realization plans, preservation traces, and answer objects.

\item It instantiates the framework through a claim-conflict answer-object family and uses it to motivate a research agenda on target-contract derivation, governed answer-object libraries, target-aware realization planning, preservation traces, and target-fidelity benchmarks.
\end{enumerate}

The rest of the paper is organized as follows. Section~\ref{sec:motivation} introduces motivating examples and the three target-status regimes. Section~\ref{sec:target-contracts} develops the formal view of environment profiles, target contracts, adequacy, realization, and preservation. Section~\ref{sec:claim-conflict} presents the claim-conflict answer-object family as a worked target. Section~\ref{sec:environment-classes} discusses environment classes and regime diagnosis. Section~\ref{sec:terrain} outlines the research agenda. Section~\ref{sec:related} positions the paper relative to natural-language interfaces, data integration, polystores, document extraction, provenance, semantic query systems, and tool-using agents.

\section{Motivating Examples and Target Status}
\label{sec:motivation}

The previous section argued that natural-language-initiated querying does not always begin with a fixed backend language. This section illustrates the distinction through three examples. The examples are not intended as a benchmark. Their purpose is to show how the semantic task changes depending on whether the target is already known, known only as an answer-object family, or not yet fixed by the environment.

\subsection{Example 1: A fixed-target query}

Consider the query:

\begin{quote}
\textit{List all suppliers in California with contracts above \$1M.}
\end{quote}

If the environment is a relational database with tables for suppliers, contracts, and locations, the target is naturally SQL. The system must identify the relevant tables, attributes, constants, predicates, and joins. It may need to resolve that ``California'' refers to a state field and that ``contracts above \$1M'' refers to a numeric threshold over contract value. These are nontrivial semantic parsing and schema-grounding problems, but the kind of target is not in question. The answer is expected to be the result of a query in a known backend language.

The same pattern holds for many graph and RDF settings. If the environment exposes a property graph with a declared graph query language, a user question may naturally target a path pattern, neighborhood query, or aggregation over graph nodes and edges. If the environment exposes an RDF graph and ontology with a SPARQL endpoint, the target may naturally be a SPARQL query. In these cases, the core task is translation into a known target family.

This is the fixed-target regime. The target language and answer form are supplied by the environment.

\subsection{Example 2: A partially specified answer-object target}

Now consider a different query:

\begin{quote}
\textit{Show how this product issue evolved across internal emails, presentations, and reports.}
\end{quote}

This query is not naturally answered by a single tuple set or ranked document list. The user is asking for an evolution object: a temporally organized account of how an issue appeared, changed, intensified, disappeared, or was reframed across artifacts. If the environment profile already declares an \textit{issue-evolution trace} as an available answer-object family, then the target family is known. The system does not need to invent the form of the answer.

However, the target is still incomplete. The system must decide which mentions belong to the issue, how to group them into stages, what transitions to represent, which artifacts serve as witnesses, and what granularity the final trace should have. Backend operations such as retrieval, extraction, clustering, temporal ordering, and summarization may all be needed, but none of them alone defines the answer object. The target is the issue-evolution trace; the execution plan is a later realization of that target.

This is the partially specified target regime. The environment already knows the answer-object family, but the particular object must still be instantiated and completed. Evaluation must therefore ask not only whether relevant documents were found, but whether the completed trace has the right structure, temporal order, witnesses, and provenance.

\subsection{Example 3: A target-construction query}

Now consider the query:

\begin{quote}
\textit{Find internal emails that contradict the claims of this announcement.}
\end{quote}

If the environment contains a public announcement, an internal email archive, metadata, keyword search, and vector retrieval, those capabilities are useful but not sufficient to define the target. A keyword query may retrieve emails containing overlapping terms. A vector search may retrieve semantically similar passages. An LLM may produce a fluent summary. But none of these native result types is, by itself, the object requested by the query.

The answer requires a different target: a claim-conflict object. Such an object must bind claim-bearing spans in the announcement, claim-bearing spans in candidate emails, an issue-alignment relation, a conflict or contradiction relation, source-span witnesses, provenance, timestamps, and uncertainty or review status. Only after this target is identified can the system decide how to realize it through retrieval, claim extraction, entity or issue alignment, contradiction scoring, metadata lookup, and answer assembly.

This is the target-construction regime. The environment contains useful sources and operators, but it does not yet provide a declared backend language or answer-object family that fully captures the user's information need. The system must first construct or select a target object before backend-specific execution can be meaningfully planned.

\subsection{Example 4: A catalog-mediated target}

A second target-construction example arises in data discovery:

\begin{quote}
\textit{Find recent county-level datasets on wildfire smoke exposure with spatial coverage, variable descriptions, and access restrictions.}
\end{quote}

Here the environment may contain multiple catalogs and metadata standards. One source may expose CKAN-style records, another schema.org metadata, another geospatial metadata, and another local domain-specific schema. The query is not merely asking for keyword matches over catalog records --- it asks for datasets satisfying several metadata commitments: topic, recency, county-level spatial coverage, variable descriptions, access conditions, and perhaps licensing or lineage.

If the environment already exposes a normalized catalog-discovery target with these fields, then the query may be partially specified: the family is known, and the system must instantiate it. If not, the system must first determine the target vocabulary that makes the query comparable across metadata standards. The target must represent dataset identity, variables, spatial coverage, temporal extent, access policy, and provenance before retrieval and ranking can be judged meaningful.

This example shows that target construction is not only a document-evidence problem. It also appears when natural-language queries range over heterogeneous metadata environments whose native schemas do not share a common answer vocabulary.

\subsection{The three target-status regimes}

These examples motivate three target-status regimes.

\noindent\textbf{Fixed-target regime.}
A query is fixed-target when the environment already supplies the target language and answer form. The system's task is to translate the natural-language query into that known family. SQL, SPARQL, Cypher, GQL, keyword search, or API calls may all serve as fixed targets when they fully capture the required answer object.

\noindent\textbf{Partially specified target regime.}
A query is partially specified when the environment already declares the relevant answer-object family, but the particular object must still be instantiated. The system must complete an object such as an evidence bundle, comparison, timeline, issue-evolution trace, or catalog-discovery target. Backend queries may be used during realization, but they do not by themselves define the answer object.

\noindent\textbf{Target-construction regime.}
A query is in the target-construction regime when the environment does not yet declare a backend language or answer-object family adequate for the user's information need. The system must first identify or construct the target object, vocabulary, or evidence structure that execution should preserve.

The regimes are relative to the environment profile, not intrinsic to the natural-language string. The same contradiction query may be target construction in an environment with only raw documents and search indexes. It may become partially specified if the environment registers a claim-conflict answer-object family and claim-extraction operators. It may become fixed-target if the environment exposes a claim graph with a declared \texttt{contradicts} relation and source-span provenance queryable through a graph language.

\begin{definition}[Target-status regimes]
\label{def:target-status}
Let $q$ be a natural-language query, $\Gamma(E)$ an environment profile, and $\tau$ an adequate target contract for $q$ in $E$. The query is \textit{fixed-target} relative to $\Gamma(E)$ if there exists a declared language $L \in \Gamma(E)$ such that every $e \in \operatorname{Obl}(\tau)$ has fit status $\textsf{exact}$ in $L$. It is \textit{partially specified} if a declared family $K \in \mathcal{A}$ covers $\operatorname{Obl}(\tau)$ but no single $L$ achieves exact fit for all obligations. It is in the \textit{target-construction} regime if neither condition holds. 
\end{definition}
These regime definitions are made precise in Definition~\ref{def:target-status} after the formal vocabulary of Section~\ref{sec:target-contracts} has been established. All three regimes are relative to the pair $(q, \Gamma(E))$: the same natural-language string may fall into different regimes under different profiles.

\subsection{Three Core Questions}

The examples separate three questions that are often conflated. The first is what the user wants returned. In fixed-target settings, the backend result type often answers this question directly; in target-construction settings, the desired answer object must be made explicit before execution can be planned. The second is what must be preserved during execution. A contradiction answer must preserve claims, conflict relations, witnesses, provenance, and uncertainty; a catalog-discovery answer must preserve metadata commitments across standards. These are not incidental output-formatting choices --- they are part of the target. The third is what operations can realize the target. Retrieval, extraction, graph queries, SQL, metadata mapping, ranking, model calls, and human review may all contribute to execution, but the execution plan should be judged against the target object it is supposed to realize.

The next section formalizes these ideas using environment profiles, target contracts, realization plans, preservation traces, and answer objects. The purpose of the formal vocabulary is not to introduce a universal intermediate language, but to make explicit what the target is, what obligations it carries, and what execution must preserve.

\section{Target Contracts, Adequacy, and Preservation}
\label{sec:target-contracts}

Section~\ref{sec:motivation} introduced three target-status regimes through examples. This section gives the minimal formal vocabulary needed to make those regimes operational. The goal is to identify the objects a target-aware system must expose in order to decide what target is adequate, how that target can be realized, and what semantic obligations must be preserved during execution.

The formal spine is:
\[
\Gamma(E)
\rightsquigarrow \tau
\rightsquigarrow \pi
\rightsquigarrow (\rho, a).
\]
Here $\Gamma(E)$ is an environment profile, $\tau$ is a target contract, $\pi$ is a realization plan, and the pair $(\rho, a)$ is the returned package: $\rho$ is the preservation trace and $a$ is the answer object. The trace and the answer object are distinct artifacts. $a$ contains the substantive bindings, relations, witnesses, provenance, and uncertainty required by $\tau$. $\rho$ records, for each obligation in $\operatorname{Obl}(\tau)$, the status with which that obligation survived realization. Separating them means that $a$ can be inspected on its own merits while $\rho$ provides the qualification layer that determines whether $a$ counts as a full or trace-qualified satisfaction of $\tau$. Each $\rightsquigarrow$ denotes a \textit{gives-rise-to} relation rather than a deterministic function: multiple valid target contracts may exist for a given profile and query, multiple realization plans may satisfy a given contract, and so on. The spine records the intended design-time and query-time progression without implying a unique derivation at each step.

\begin{figure*}[t]
\centering
\small
\fbox{%
\begin{minipage}{0.82\textwidth}
\centering
\textbf{NL question}
\\[2pt]
$\downarrow$
\\[2pt]
\textbf{target-contract generator}
\\[2pt]
$\downarrow$
\\[2pt]
\textbf{contract obligations}
\\[2pt]
$\downarrow$
\\[2pt]
\textbf{heterogeneous retrieval planner}
\\[2pt]
$\downarrow$
\\[2pt]
\textbf{operator selection + preservation annotations}
\\[2pt]
$\downarrow$
\\[2pt]
\textbf{backend execution}
\\[2pt]
$\downarrow$
\\[2pt]
\textbf{runtime lineage + preservation trace}
\\[2pt]
$\downarrow$
\\[2pt]
\textbf{answer object + trace-qualified certification}
\end{minipage}%
}
\caption{Operational flow of target-aware natural-language querying over a heterogeneous backend. The preservation trace begins during planning as obligation-level annotations attached to operator choices, and is instantiated during execution using runtime lineage, provenance, returned artifacts, scores, and missing values.}
\label{fig:target-aware-flow}
\end{figure*}

\subsection{Environment Profiles}

Let $E$ denote a data environment. In a fixed-target setting, $E$ may be summarized by a schema and a backend language. In the heterogeneous settings considered here, $E$ may include structured records, graphs, document collections, metadata catalogs, search indexes, extraction layers, entity-linking resources, provenance stores, and workflow tools. For target formation, what matters is not only what data are present, but what capabilities the environment exposes.

\noindent\textbf{Definition 1. Environment profile.}
An environment profile is a tuple
\[
\Gamma(E)=
\langle
\mathcal{S},
\mathcal{M},
\mathcal{O},
\mathcal{A},
\mathcal{P}
\rangle,
\]
where $\mathcal{S}$ is a set of sources and source descriptions, $\mathcal{M}$ is a set of mappings or alignments, $\mathcal{O}$ is a set of available operators, $\mathcal{A}$ is a set of declared answer-object families, and $\mathcal{P}$ is a set of provenance and uncertainty mechanisms.

The profile is not a \textit{mediated schema} because a mediated schema provides a common logical surface over sources. By contrast, $\Gamma(E)$ is a capability description. It records what sources exist, how they are semantically related, what operations can be invoked, what answer-object families are already known to the system, and what provenance or uncertainty information can be carried forward.

\subsection{Target Contracts}

A natural-language query expresses an information need. A target contract specifies what kind of answer object would be adequate for that need in a given environment.

\begin{definition}[Target contract]
\label{def:target-contract}
A target contract is a tuple
\[
\tau =
\langle
K,
B,
R,
W,
P,
U,
C
\rangle,
\]
where $K$ is an answer-object kind, $B$ is a set of required bindings or slots, $R$ is a set of required relations among those bindings, $W$ is a set of witness obligations, $P$ is a set of provenance obligations, $U$ is a set of uncertainty obligations, and $C$ is a set of admissibility constraints.    
\end{definition}

\begin{definition}[Obligation set]
\label{def:obligation-set}
The \textit{obligation set} of a target contract
$\tau = \langle K, B, R, W, P, U, C \rangle$ is
\[
\operatorname{Obl}(\tau) = B \cup R \cup W \cup P \cup U.
\]
The admissibility constraints $C$ are excluded from $\operatorname{Obl}(\tau)$: they govern what counts as an admissible answer object but do not impose independent realization obligations on the execution plan. The answer-object kind $K$ similarly identifies the family to be instantiated rather than an obligation to be discharged by a specific operator.
\end{definition}

The answer-object kind $K$ identifies the family of object to be constructed, such as a tuple set, ranked list, evidence bundle, comparison object, timeline, claim-conflict object, catalog-discovery object, or workflow-backed report. The bindings $B$ specify the typed entities or values that must be bound. The relations $R$ specify how those bindings must be connected. Witness obligations $W$ specify what evidence must be attached to justify the answer. Provenance obligations $P$ specify what source, span, version, or derivation information must be retained. Uncertainty obligations $U$ specify what scores, thresholds, confidence annotations, abstentions, or review flags must accompany the answer. Constraints $C$ specify type restrictions, source restrictions, admissible operator classes, threshold rules, or consistency conditions. A target contract is therefore not an executable query. It is a specification of what the answer object must contain and what execution must preserve.

\begin{definition}[Contract satisfaction]
\label{def:contract-satisfaction}
Let $(\rho, a)$ be a returned package. We write $a \models \tau$ when the answer object $a$ satisfies the target contract $\tau$: the required bindings are instantiated or explicitly marked as unavailable, the required relations are represented, the required witnesses and provenance are attached, the required uncertainty information is present, and the admissibility constraints are satisfied. We write $(\rho, a) \models_{\mathsf{tq}} \tau$ (trace-qualified satisfaction) when every obligation in $\operatorname{Obl}(\tau)$ with $\rho(e) \neq \textsf{preserved}$ is explicitly recorded in $\rho$ and $a$ does not present those obligations as fully discharged.
\end{definition}
This satisfaction relation is contract-relative. A ranked list of documents may satisfy a retrieval contract but fail a claim-conflict contract if it does not identify the relevant claims, align their issues, represent the conflict relation, attach source-span witnesses, and record uncertainty.

\subsection{Design-Time Workload Adequacy Profiling}
\label{sec:workload-adequacy}

Fixed-target systems can assume that the target schema and answer type are given. In \nlx they are not: the system designer must analyze whether the environment can support the target objects a workload requires. This analysis is a design-time activity called \textit{workload adequacy profiling}.

Let $\mathcal{W}={q_1,\ldots,q_m}$ be a representative workload for environment $E$. For each $q_i$, the designer derives a candidate contract $\tau_i$ using the structure of Definition~2. The result is a workload--environment adequacy profile: for each query, its candidate contract, its regime classification relative to $\Gamma(E)$, and the fit status of each obligation. This profile tells the system builder which workload regions are supported, which require new declared families or extraction layers, and which cannot be supported without constructing new targets.

\noindent\textbf{Operational procedure.}
A minimal design-time procedure has five steps.
\begin{enumerate}
\item \textbf{Choose a representative workload.}
The designer first identifies a set of natural-language questions that represent the intended use of the environment. The workload should include not only simple lookup queries, but also the complex cases that motivate \nlx: comparisons, conflicts, timelines, catalog-discovery tasks, evidence bundles, cross-source witnesses, workflow-backed reports, and other answer objects not naturally represented as a single backend query.

\item \textbf{Derive candidate commitments for each query.}
For each $q_i \in \mathcal{W}$, the system derives a candidate contract $\tau_i$. This step is weaker than full semantic parsing. Its purpose is to expose the answer obligations proposed by the query-analysis step: the requested answer kind, candidate bindings, candidate relations, witness expectations, provenance expectations, uncertainty expectations, and admissibility constraints.

\item \textbf{Normalize recurring commitments into candidate answer families.}
Across the workload, many queries will induce similar commitments. Several queries may require claim-conflict objects, evidence timelines, dataset-discovery records, comparison objects, or cross-source witness bundles. The designer groups recurring contracts into candidate answer-object families. This step is the \nlx analogue of schema design: instead of assuming a fixed target schema, the designer discovers which target families the workload requires.

\item \textbf{Check obligation-level support against the environment profile.}
For each obligation $e \in \operatorname{Obl}(\tau_i)$, the designer or compiler checks whether $\Gamma(E)$ exposes sources, mappings, operators, indexes, extraction layers, provenance mechanisms, or review procedures capable of supporting that obligation. This produces a fit status
\begin{quote}
$\operatorname{fit}(e,\Gamma(E)) \in \{\textsf{exact}, \textsf{approximate}, \textsf{weakened}, \textsf{review}, \bot\}$
\end{quote}
The value $\textsf{exact}$ means that the environment can support the obligation directly. The value $\textsf{approximate}$ means that the obligation can be attempted through a model, score, heuristic, similarity function, or thresholded operator. The value $\textsf{weakened}$ means that only a lower-granularity or less precise version is available. The value $\textsf{review}$ means that the system can produce a candidate but cannot safely discharge the obligation without human or external validation. The value $\bot$ means that the environment exposes no known support for the obligation.

\item \textbf{Diagnose adequacy gaps and environment repairs.}
The final step summarizes the workload regions that are supported, partially supported, or unsupported. The diagnosis may recommend adding mappings, declaring new answer-object families, building extraction layers, adding span-level provenance, registering new operators, enriching metadata, or restricting the advertised workload. The output is more than a classification of queries --- it is a design plan for making the environment more adequate for the intended workload.
\end{enumerate}

\noindent\textbf{Example.}
Consider the workload query:
\begin{quote}
\textit{Find internal emails that contradict the claims of this announcement.}
\end{quote}
A candidate-commitment extraction would propose $K_i=\textsc{ClaimConflict}$. The candidate bindings would include the announcement, announcement claims, candidate emails, email claims, issues, and source spans. The candidate relations would include \texttt{containsSpan}, \texttt{sameIssue}, and \texttt{contradicts}. The witness and provenance obligations would include paired source-span witnesses, document identifiers, timestamps, span offsets, and extraction versions. The uncertainty obligations would include confidence or review status for issue alignment and contradiction detection. The admissibility constraints would restrict candidate artifacts to internal emails and require low-confidence contradictions to be returned as review-dependent candidates.

The design-time question is then not simply how to answer this query, but whether the environment has the target resources needed: announcement parsing, email access, claim extraction, issue alignment, contradiction detection, span provenance, document timestamps, and review annotation. If the environment has only keyword search over emails, the query cannot be treated as a fully supported target. If it has email search and claim extraction but only document-level provenance, the workload may be supported only with weakened witness obligations. If it already declares a \textsc{ClaimConflict} answer family with the required bindings, relations, and witness fields, the query is partially specified rather than target-construction.

\noindent\textbf{Workload-Level Regime Diagnosis.} The three target-status regimes are best understood as outcomes of the workload adequacy profile, not only as classifications of individual queries.

\begin{definition}[Target-status regimes]
\label{def:target-status-regime}
Let $q$ be a natural-language query, $\Gamma(E)$ an environment profile, and $\tau$ an adequate target contract for $q$ in $E$. The query is \textit{fixed-target} relative to $\Gamma(E)$ if there exists a declared language $L \in \Gamma(E)$ such that every
$e \in \operatorname{Obl}(\tau)$ has fit status $\textsf{exact}$ in $L$. It is \textit{partially specified} if a declared family $K \in \mathcal{A}$ covers $\operatorname{Obl}(\tau)$ but no single $L$ achieves exact fit for all obligations.
It is in the \textit{target-construction} regime if neither condition holds. All three regimes are relative to the pair $(q, \Gamma(E))$.
\end{definition}

This diagnosis is relative to both the workload and the environment. The same natural-language query may move from target construction to partial specification, or from partial specification to fixed-target translation, as the environment gains new mappings, extraction layers, answer-object families, or queryable semantic relations. Conversely, expanding the workload may reveal new inadequacies: adding a query pattern that requires span-level witnesses, temporal supersession, or contradiction review may move part of the workload outside the current adequacy frontier.

\subsection{Query-Time Use of the Design-Time Profile}
\label{sec:query-time-profile-use}

Although workload adequacy profiling is a design-time activity, its results are used at query time. When a new query arrives, the system can compare its candidate commitments against the previously profiled workload regions. If the query matches a supported fixed-target region, translation may proceed. If it matches a declared answer-object family, the system can instantiate that family and plan realization. If it falls outside the profiled regions, the system can either trigger target construction, ask for clarification, return a trace-qualified partial answer, or report that the environment lacks the capability to support the requested target.

The design-time analysis does not replace runtime query understanding. It constrains and informs it. The additional burden introduced by \nlx is that the target space itself must be profiled, designed, and evolved from the natural-language query workload.

\subsection{Realization Plans}

Once a target contract has been selected or constructed, the system must attempt to realize it using the capabilities of the environment.

\noindent\textbf{Definition 4. Realization plan.}
A realization plan for target contract $\tau$ in environment $E$ is a sequence or directed acyclic graph of operators $\pi = \langle o_1,\ldots,o_n\rangle$, where each $o_i \in \mathcal{O}$ is available in $\Gamma(E)$. We write $\operatorname{Realizes}(\pi,\tau,\Gamma(E))$ when executing $\pi$ attempts to construct an answer object satisfying $\tau$.

A realization plan may include conventional queries, retrieval calls, extraction operators, graph traversals, entity-resolution steps, metadata mappings, ranking operations, model calls, or human-review steps. The plan is distinct from the target: the target contract states what must be produced and preserved; the plan states how the system attempts to produce it. This distinction matters because two plans may realize the same target contract in different environments, and two plans that invoke similar tools may realize different targets if they preserve different obligations.

\subsection{Preservation Traces}

Realization can degrade target obligations. A contract may require source-span evidence, while an available source provides only document-level provenance. A contract may require semantic contradiction, while the environment provides only a thresholded textual-entailment score. A contract may require entity identity, while the system can only compute approximate entity matches. These cases should not be hidden inside the execution plan --- they should be recorded as part of the answer.

\begin{definition}[Preservation trace]
\label{def:preservation-trace}
A preservation trace for target contract $\tau$ is a total function
\[
\rho : \operatorname{Obl}(\tau) \rightarrow \mathcal{D},
\]
where
\[
\begin{aligned}
\mathcal{D} = \{
\textsf{preserved},\;
\textsf{approximated},\;
\textsf{weakened},\;\\
\textsf{requiresReview},\;
\textsf{unsupported}
\}
\end{aligned}
\]
The status \textsf{preserved} means that the obligation is carried through realization without semantic weakening. The status \textsf{approximated} means that the obligation is realized using an approximate proxy, model, threshold, score, or heuristic, but the type of the returned artifact matches what the contract requires. The status \textsf{weakened} means that the type of the returned artifact is itself coarser than what the contract requires, so the obligation is relaxed rather than merely approximated. The status \textsf{requiresReview} means that the obligation cannot be safely discharged automatically. The status \textsf{unsupported} means that $\Gamma(E)$ exposes no operator or source capable of even partially addressing the obligation; this corresponds to the $\bot$ outcome of the design-time fit test. An answer object whose trace contains any \textsf{unsupported} entry cannot be returned as even a trace-qualified realization: the system must report an environment gap instead.
\end{definition}
\begin{table}[t]
\centering
\small
\begin{tabular}{|l|l|p{3.3cm}|}
\hline
\textbf{Design-time fit} & \textbf{Runtime trace status} & \textbf{Meaning} \\
\hline
\textsf{exact}       & \textsf{preserved}       & Capability exists; obligation discharged without weakening. \\
\hline
\textsf{approximate} & \textsf{approximated}    & Proxy available; correct type returned by imprecise method. \\
\hline
\textsf{weakened}    & \textsf{weakened}        & Only coarser-type capability available; type precision relaxed. \\
\hline
\textsf{review}      & \textsf{requiresReview}  & Automatic discharge unsafe; human or external validation needed. \\
\hline
$\bot$               & \textsf{unsupported}     & No capability exists; trace-qualified return not permitted. \\
\hline
\end{tabular}
\caption{Correspondence between design-time fit values and runtime
preservation trace statuses. The fit value is a planning-time prediction;
the trace status is its runtime instantiation.}
\label{tab:fit-trace-correspondence}
\end{table}

\noindent\textbf{Correspondence between design-time fit and runtime trace statuses.}
The fit function $\operatorname{fit}(e,\Gamma(E))$ used in design-time workload profiling and the preservation trace domain $\mathcal{D}$ used at runtime share four of their five values, but serve distinct roles. The design-time value $\textsf{exact}$ asserts that the environment exposes a capability that can support obligation $e$ without semantic weakening; at runtime, successful exercise of that capability produces trace status $\textsf{preserved}$. The design-time value $\textsf{approximate}$ asserts that only an approximate proxy is available; at runtime this produces $\textsf{approximated}$. The design-time value $\textsf{weakened}$ asserts that only a coarser-type capability is available; at runtime this produces $\textsf{weakened}$.
The design-time value $\textsf{review}$ asserts that automatic discharge is unsafe; at runtime this produces $\textsf{requiresReview}$. The design-time value $\bot$ asserts that no capability exists at all; at runtime this produces $\textsf{unsupported}$, which as noted above prevents even trace-qualified certification. The design-time fit value is therefore a prediction and a planning constraint; the runtime trace status is its instantiation against concrete artifacts, scores, and source responses. A design-time fit of $\textsf{exact}$ does not guarantee a runtime status of $\textsf{preserved}$: a source declared as supporting span offsets may fail to return them for a particular document, in which case the runtime status degrades to $\textsf{weakened}$ even though the design-time fit was $\textsf{exact}$. This is summarized in Table \ref{tab:fit-trace-correspondence}.

\noindent\textbf{Remark.} The distinction between \textsf{approximated} and \textsf{weakened} turns on the \textit{type} of the returned artifact relative to the obligation. If the contract requires span-level provenance and the operator returns span-level provenance produced by a scoring heuristic, the obligation is \textsf{approximated}: the right type is present, but the method is imprecise. If the operator returns only document-level provenance, the obligation is \textsf{weakened}: the type itself is coarser than the contract requires. Approximation is an instrument-quality issue; weakening is a type-precision issue.

The elements introduced during realization --- extracted spans, derived confidence scores, intermediate clusters, candidate alignments --- are treated separately from preservation status. Such elements are admissible only if they do not violate the constraints in $C$ and are exposed when they affect the final answer object.

Figure~\ref{fig:target-aware-flow} shows where preservation traces enter the architecture. The trace begins during realization planning, when the compiler matches target obligations to operators with declared capability signatures. It is then instantiated during execution using the actual artifacts, lineage, provenance, scores, and missing values returned by the plan. The preservation trace differs from an execution log: an execution log records what the system did, while a preservation trace records what happened to the obligations of the target contract.

Operator capability signatures support this two-level process. A useful decomposition distinguishes a \textit{static capability signature} from a \textit{runtime trace resolution}. An operator $o$ declares at design time what obligation types it can address and the best and worst trace status it may achieve:
\[
\sigma(o):
\mathsf{ObligationType} \mapsto
[\mathsf{best},\,\mathsf{worst}],
\]
where $\mathsf{best}$ and $\mathsf{worst}$ are elements of $\mathcal{D}$. The planner uses this declaration at planning time: an operator that can achieve at best \textsf{approximated} for a given obligation type will never be assigned to discharge that obligation as \textsf{preserved}.

The actual trace status for a specific obligation is resolved at execution time from the artifacts the operator returns. A vector-retrieval operator, for example, declares
\[
\sigma(o_{\mathsf{vecRet}}):
\mathsf{CandidateBinding} \mapsto
[\textsf{approximated},\,\textsf{approximated}],
\]
advertising that it always achieves exactly \textsf{approximated} for candidate-binding obligations. A provenance operator may declare
\[
\sigma(o_{\mathsf{prov}}):
\mathsf{SpanProvenance} \mapsto
[\textsf{preserved},\,\textsf{weakened}],
\]
advertising that it achieves between \textsf{weakened} and \textsf{preserved} depending on what the source returns. At runtime, the trace is resolved as:
\[
\rho(\mathsf{SpanProvenance}) =
\begin{cases}
\textsf{preserved} & \text{if source offsets are returned},\\
\textsf{weakened} & \text{if only document IDs are returned}.
\end{cases}
\]
The static signature constrains the planner; the runtime resolution instantiates the trace. Both are necessary: the planner needs the static declaration to attach preliminary obligation-level annotations, while the runtime resolution is needed because some conditions --- such as whether a source stores span offsets --- cannot be known until execution.

\subsection{Certification}
\label{sec:no-unreported-degradation}

Preservation traces support the central certification condition needed for target-aware querying. The condition has two roles. At design time, it is an invariant used to evaluate whether an environment can support a representative workload without hiding target degradation. At query time, it is applied to a concrete answer object after realization.

\noindent\textbf{Certification condition: no unreported degradation.}
Let $\tau$ be a target contract, $\pi$ a realization plan, $\rho$ its preservation trace, and $a$ the answer object produced by executing $\pi$. If an obligation $e \in \operatorname{Obl}(\tau)$ is approximated, weakened, or deferred to review during realization, then $a$ should not be certified as fully satisfying $\tau$. It may instead be returned as a trace-qualified realization of $\tau$, provided that $\rho(e)$ records the corresponding status.

At design time, this condition constrains workload adequacy profiling. An environment should not be described as fully adequate for a query pattern if the obligations induced by that pattern can only be supported approximately, at lower granularity, or through human review. Such a workload region may still be useful, but it should be classified as trace-qualified, weakened, review-dependent, or outside the current adequacy frontier.

At query time, the same condition is enforced on the realized answer. The system compares the obligations of $\tau$ with the statuses recorded in $\rho$. If all required obligations are preserved, the answer may be certified as satisfying the target contract. If some obligations are approximated, weakened, or marked as requiring review, the answer may still be returned, but only as a trace-qualified realization. For example, if a claim-conflict object relies on approximate issue alignment, the trace should say so. If a provenance obligation is weakened from span-level to document-level provenance, the trace should say so. If a conflict relation requires human validation, the trace should mark the relevant obligation as requiring review.

This condition does not claim the correctness of realizations or operators. It states a weaker but essential requirement: the system must not present a degraded realization as if it had fully satisfied the original target contract. Certification therefore connects the design-time adequacy profile with the query-time answer object, making visible whether the semantic obligations of the target survived realization.

\subsection{Implications for Evaluation}

The objects defined above induce a target-fidelity evaluation problem. A target-aware system can fail at several levels: choosing the wrong answer-object kind $K$; omitting required bindings or relations; returning claims without witnesses; dropping provenance; hiding uncertainty; weakening a target obligation without recording the weakening; or producing an answer object that appears plausible but does not satisfy the target contract.

For a benchmark query $q$, evaluation should therefore ask at least five questions. Did the system select the correct target-status regime? Did it derive or select an adequate contract $\tau$? Did the realization plan $\pi$ cover the required obligations of the contract? Did the preservation trace $\rho$ correctly record preserved, approximated, weakened, and review-dependent obligations? Did the final answer object $a$ satisfy the contract, or was it appropriately returned as a trace-qualified realization?

The next section instantiates these ideas through the claim-conflict answer-object family.

\section{Answer-Object Family: The Claim-Conflict Case}
\label{sec:claim-conflict}

This section instantiates the formal vocabulary from Section~\ref{sec:target-contracts} through a single answer-object family: \textsc{ClaimConflict}. The purpose is to show how a natural-language query that appears to ask for documents actually requires a structured target object with bindings, relations, witnesses, provenance, uncertainty, and preservation status.

\subsection{Query and Environment Fragment}

Consider the query:

\begin{quote}
\textit{Find internal emails that contradict the claims of this announcement.}
\end{quote}

Assume an environment $E_{\mathsf{cc}}$ containing a public-announcement collection, an internal-email archive, document metadata, a text index, a vector index, a claim-span extractor, an issue-alignment model, a contradiction-scoring model, and a provenance store. A simplified environment profile is:
\[
\Gamma(E_{\mathsf{cc}})=
\langle
\mathcal{S}_{\mathsf{cc}},
\mathcal{M}_{\mathsf{cc}},
\mathcal{O}_{\mathsf{cc}},
\mathcal{A}_{\mathsf{cc}},
\mathcal{P}_{\mathsf{cc}}
\rangle.
\]
Here $\mathcal{S}_{\mathsf{cc}}$ contains the announcement collection, the email archive, and metadata records. The mappings $\mathcal{M}_{\mathsf{cc}}$ include document-to-metadata links, actor-name normalization, and tentative issue alignments. The operator set $\mathcal{O}_{\mathsf{cc}}$ includes keyword retrieval, vector retrieval, claim extraction, issue alignment, contradiction scoring, metadata lookup, and answer assembly. The answer-object family set $\mathcal{A}_{\mathsf{cc}}$ contains \textsc{ClaimConflict} if the environment has already registered this target family. The provenance and uncertainty mechanisms $\mathcal{P}_{\mathsf{cc}}$ include document identifiers, span offsets, timestamps, extraction versions, model versions, confidence scores, and review flags.

The native outputs of the available backends are not sufficient by themselves. Keyword search returns documents. Vector search returns passages. A contradiction model returns scores. Metadata lookup returns identifiers and timestamps. The user's query, however, asks for an answer object that explains which announcement claim conflicts with which internal-email claim, and why that conflict is supported by inspectable evidence.

\subsection{Target Contract}

The target contract for this query has answer-object kind $K=\textsc{ClaimConflict}$. A compact version is:
\[
\tau_{\mathsf{cc}} =
\langle
K_{\mathsf{cc}},
B_{\mathsf{cc}},
R_{\mathsf{cc}},
W_{\mathsf{cc}},
P_{\mathsf{cc}},
U_{\mathsf{cc}},
C_{\mathsf{cc}}
\rangle.
\]
The bindings $B_{\mathsf{cc}}$ include a reference artifact, a reference claim span, a candidate email, a candidate claim span, an issue object, timestamps, and source identifiers. The relations $R_{\mathsf{cc}}$ include \texttt{containsSpan}, \texttt{sameIssue}, and \texttt{contradicts}. The witness obligations $W_{\mathsf{cc}}$ require paired source spans: one from the announcement and one from the candidate email. The provenance obligations $P_{\mathsf{cc}}$ require document identifiers, span offsets, timestamps, extraction versions, and model or operator versions. The uncertainty obligations $U_{\mathsf{cc}}$ require confidence scores or review flags for issue alignment and contradiction. The constraints $C_{\mathsf{cc}}$ specify when the system may assert a contradiction and when it must return a review-dependent candidate. Table~\ref{tab:claim-conflict-contract-new} summarizes the contract.

\begin{table*}[t]
\centering
\small
\begin{tabular}{|p{3.1cm}|p{5.2cm}|p{7.2cm}|}
\hline
\textbf{Contract component} & \textbf{Claim-conflict instance} & \textbf{Purpose} \\
\hline
$K$ &
\textsc{ClaimConflict} &
Declares that the answer is a conflict object, not merely a document list or generated summary. \\
\hline
$B$ &
reference artifact, reference claim span, candidate email, candidate claim span, issue, timestamp, source identifier &
Declares the typed slots that must be bound before a conflict answer can be returned. \\
\hline
$R$ &
\texttt{containsSpan}, \texttt{sameIssue}, \texttt{contradicts} &
Declares the relations that make the answer a conflict object rather than a set of related documents. \\
\hline
$W$ &
paired source spans and supporting metadata &
Requires inspectable evidence for both sides of the conflict. \\
\hline
$P$ &
document IDs, span offsets, timestamps, corpus links, extraction versions, model versions &
Records where the evidence came from and how the conflict object was derived. \\
\hline
$U$ &
issue-alignment score, contradiction score, threshold, review flag &
Distinguishes exact anchors from approximate or review-dependent judgments. \\
\hline
$C$ &
type constraints, source restrictions, threshold and review rules &
Specifies admissible answers; for example, low-confidence contradiction scores may be returned only as review-dependent candidates. \\
\hline
\end{tabular}
\caption{A target contract for the \textsc{ClaimConflict} answer-object family.}
\label{tab:claim-conflict-contract-new}
\end{table*}

This target contract is different from an execution plan. It does not say whether the candidate emails should be found using keyword search, vector search, graph expansion, or metadata filters. It states what must be present in the answer object for the result to count as a claim-conflict answer.

\subsection{Contract Fit}

The contract-fit test asks how each obligation in $\operatorname{Obl}(\tau_{\mathsf{cc}})$ is supported by $\Gamma(E_{\mathsf{cc}})$. In this example, some obligations are exact, some are approximate, and some may require review:
\[
\operatorname{fit}(e,\Gamma(E_{\mathsf{cc}})) \in
\{
\textsf{exact},\;
\textsf{approximate},\;
\textsf{weakened},\;
\textsf{review},\;
\bot
\}.
\]
Stable document identifiers have exact fit if the announcement and emails have persistent IDs. Source-span witnesses have exact fit if the environment stores offsets for extracted spans; they have weakened fit if only document-level links are available. Issue alignment usually has approximate fit because it depends on entity matching, topic similarity, or embedding-based alignment. The contradiction relation may have approximate fit if it is scored by a model, and review fit if the score is below the threshold required for automatic assertion. Table~\ref{tab:claim-conflict-fit} illustrates the fit assessment.

\begin{table}[t]
\centering
\small
\begin{tabular}{|p{3.2cm}|p{3.9cm}|}
\hline
\textbf{Obligation} & \textbf{Fit in $\Gamma(E_{\mathsf{cc}})$} \\
\hline
Bind announcement artifact & \textsf{exact}: supplied by query context or document ID \\
\hline
Bind candidate emails & \textsf{approximate}: retrieved by keyword/vector search \\
\hline
Bind claim spans & \textsf{approximate}: produced by claim-span extractor \\
\hline
Preserve document IDs & \textsf{exact}: stored in metadata \\
\hline
Preserve span offsets & \textsf{exact} if extractor records offsets; otherwise \textsf{weakened} \\
\hline
Establish same issue & \textsf{approximate}: entity/topic/embedding alignment \\
\hline
Assert contradiction & \textsf{approximate} or \textsf{review}: model score and threshold \\
\hline
Attach timestamps & \textsf{exact}: metadata lookup \\
\hline
\end{tabular}
\caption{Illustrative contract-fit assessment for the claim-conflict query.}
\label{tab:claim-conflict-fit}
\end{table}

This fit table determines the target-status regime relative to the profile. If \textsc{ClaimConflict} is not registered in $\mathcal{A}_{\mathsf{cc}}$, then the query requires target construction. If it is registered but the object must be populated through retrieval, extraction, alignment, and scoring, then the query is in the partially specified regime. If the environment already exposes a claim graph with queryable \texttt{contradicts} edges and span provenance, then the same query may become fixed-target over the graph language.

\subsection{Realization Plan}

A realization plan $\pi_{\mathsf{cc}}$ attempts to construct an answer object satisfying $\tau_{\mathsf{cc}}$. One possible plan is:
\[
\pi_{\mathsf{cc}} =
\langle
o_{\mathsf{ref}},
o_{\mathsf{ret}},
o_{\mathsf{claim}},
o_{\mathsf{align}},
o_{\mathsf{contr}},
o_{\mathsf{prov}},
o_{\mathsf{asm}}
\rangle.
\]
The operator $o_{\mathsf{ref}}$ identifies claim-bearing spans in the reference announcement. The retrieval operator $o_{\mathsf{ret}}$ finds candidate internal emails. The claim extractor $o_{\mathsf{claim}}$ identifies claim-bearing spans in candidate emails. The alignment operator $o_{\mathsf{align}}$ estimates whether a reference claim and candidate claim concern the same issue. The contradiction operator $o_{\mathsf{contr}}$ estimates whether the paired claims conflict. The provenance operator $o_{\mathsf{prov}}$ attaches source identifiers, offsets, timestamps, and operator versions. The assembler $o_{\mathsf{asm}}$ constructs the final claim-conflict object.

This plan is one possible realization, not part of the target definition. A different environment might realize the same contract through a graph query over a claim graph plus a provenance lookup. Another might use a human-review queue for the contradiction relation. The target contract remains the object against which those plans are judged.

\subsection{Preservation Trace}

The preservation trace records what happened to each contract obligation during realization. Because all obligations in this example have at least approximate support, the trace ranges over $\mathcal{D} \setminus \{\textsf{unsupported}\}$:
\[
\rho_{\mathsf{cc}} :
\operatorname{Obl}(\tau_{\mathsf{cc}})
\rightarrow
\mathcal{D}.
\]
A typical trace for the plan above might mark document identifiers and timestamps as \textsf{preserved}; issue alignment as \textsf{approximated}; contradiction as \textsf{requiresReview} when the score is below an assertion threshold; and span provenance as \textsf{weakened} if offsets are unavailable for some extracted claims.

\begin{table*}[t]
\centering
\small
\begin{tabular}{|p{3.2cm}|p{4.7cm}|p{3.0cm}|p{4.3cm}|}
\hline
\textbf{Contract obligation} & \textbf{Realization step} & \textbf{Trace status} & \textbf{Explanation} \\
\hline
Reference announcement ID & Context binding / metadata lookup & \textsf{preserved} & The identifier is copied from stable source metadata. \\
\hline
Reference claim span & Claim extraction over announcement & \textsf{preserved} if offset anchored; otherwise \textsf{weakened} & The span is introduced by extraction but preserved only if anchored to source offsets. \\
\hline
Candidate email & Keyword/vector retrieval & \textsf{approximated} & Retrieval produces candidates rather than an exhaustive logical match. \\
\hline
Candidate claim span & Claim extraction over candidate email & \textsf{preserved} if offset anchored; otherwise \textsf{weakened} & The claim span must be linked to the source document. \\
\hline
Same issue relation & Entity/topic/embedding alignment & \textsf{approximated} & Alignment is score-based. \\
\hline
Contradiction relation & Contradiction scoring model & \textsf{approximated} or \textsf{requiresReview} & The relation is model-derived and may require review below an assertion threshold. \\
\hline
Source-span witnesses & Span resolver and provenance lookup & \textsf{preserved} or \textsf{weakened} & Preserved when span offsets are available; weakened when only document-level provenance exists. \\
\hline
Timestamp and corpus provenance & Metadata lookup & \textsf{preserved} & Metadata values are copied from the source records. \\
\hline
\end{tabular}
\caption{A preservation trace for the claim-conflict realization. Introduced elements, such as extracted spans or model scores, are not themselves preservation statuses; the trace records whether the corresponding contract obligations are preserved, approximated, weakened, or require review.}
\label{tab:claim-conflict-trace}
\end{table*}

The trace is also where the no-unreported-degradation condition applies. If the contradiction relation is model-derived and below the automatic assertion threshold, the returned answer object should not be certified as a fully satisfied claim-conflict object. It should instead be returned as a review-dependent realization of the contract.

\subsection{Returned Answer Object}

A returned answer object $a_{\mathsf{cc}}$ should expose the target structure, not only a textual answer or a ranked list of emails:
\[
\begin{aligned}
a_{\mathsf{cc}} =
\langle
c_r, c_e, d_r, d_e, s_r, s_e,
\mathsf{issue}, \mathsf{relation},\\
\mathsf{witnesses}, \mathsf{provenance},
\mathsf{uncertainty}
\rangle,
\end{aligned}
\]
with companion preservation trace $\rho_{\mathsf{cc}}$ returned alongside $a_{\mathsf{cc}}$ as the pair $(\rho_{\mathsf{cc}}, a_{\mathsf{cc}})$. Here $c_r$ is the reference claim from the announcement, $c_e$ is the candidate claim from the internal email, $d_r$ and $d_e$ are the corresponding documents, $s_r$ and $s_e$ are span anchors, $\mathsf{issue}$ is the aligned issue, $\mathsf{relation}$ is the asserted or candidate conflict relation, $\mathsf{witnesses}$ contains the paired evidence spans, $\mathsf{provenance}$ records source and derivation information, and $\mathsf{uncertainty}$ records scores and review status.

\noindent\textbf{Illustrative returned object.}
Consider a returned candidate conflict in which the reference claim $c_r$ from the announcement is: ``No elevated risk was observed in the tested population.'' The candidate claim $c_e$ from an internal email is: ``Subgroup testing showed elevated risk in exposed subjects.'' The aligned issue is \textit{product safety risk}, and the candidate relation is \texttt{contradicts}. The witnesses are the announcement span \texttt{doc A, offsets 122--178} and the email span \texttt{doc E17, offsets 310--380}. The uncertainty record contains a same-issue score of $0.91$ and a contradiction score of $0.78$; because the contradiction score is below the threshold for automatic assertion, the object is marked as requiring review. The preservation trace records that document identifiers were preserved, span provenance was preserved, issue alignment was approximated, and the contradiction relation requires review.

This instance fills the tuple above with concrete values, and it should not be certified as a fully satisfied \textsc{ClaimConflict} answer. It is a trace-qualified candidate conflict: useful for inspection, but explicitly marked as depending on an approximate alignment and a review-dependent contradiction judgment. The important distinction is that the system has not merely returned a relevant email or a fluent summary. It has returned a structured answer object whose required fields, supporting evidence, uncertainty, and preservation status are visible.

\subsection{The Role of Artifacts}

The claim-conflict example illustrates the role of each formal artifact. The environment profile $\Gamma(E_{\mathsf{cc}})$ determines the regime: the same natural-language query may require target construction in a raw document environment, partial specification in an environment with a declared \textsc{ClaimConflict} family, or fixed-target translation in an environment with a queryable claim graph. The target contract $\tau_{\mathsf{cc}}$ defines what the answer must contain, preventing the system from treating retrieval results, contradiction scores, or generated summaries as sufficient by themselves. The realization plan $\pi_{\mathsf{cc}}$ separates target definition from execution: the same contract can be realized through different operator plans in different environments. The preservation trace $\rho_{\mathsf{cc}}$ makes degradation explicit, so that approximate issue alignment, weakened provenance, and review-dependent contradiction are not hidden behind the final answer. Finally, the answer object $a_{\mathsf{cc}}$ gives evaluation a concrete structure: the system can be judged by whether it selected the right target family, filled the required slots, represented the required relations, attached witnesses and provenance, exposed uncertainty, and correctly trace-qualified any approximation or weakening.

This is the general pattern for \nlx systems. The target contract defines the answer object to be formed; realization constructs it; preservation traces record what survived; and evaluation asks whether the returned object satisfies, or explicitly qualifies, the target.

\subsection{Answer-Object Family: The Catalog-Discovery Case}
\label{sec:catalog-discovery-sketch}

To confirm that the \textsc{ClaimConflict} walkthrough is not idiosyncratic, we sketch a second answer-object family: \textsc{CatalogDiscovery}.
The motivating query is:
\begin{quote}
\textit{Find recent county-level datasets on wildfire smoke exposure
with spatial coverage, variable descriptions, and access restrictions.}
\end{quote}
The target contract has answer-object kind $K = \textsc{CatalogDiscovery}$. The required bindings $B$ include a dataset identifier, a topic or subject descriptor, a spatial-coverage specification, a temporal-extent record, a set of variable descriptions, and an access or license record. The required relations $R$ include \texttt{hasVariable}, \texttt{hasSpatialCoverage}, and \texttt{hasAccessPolicy}. The witness obligations $W$ require a provenance pointer to the originating catalog record for each field, so that conflicting values across metadata standards can be detected and reported. The provenance obligations $P$ require the catalog source (e.g.\ CKAN, schema.org, DCAT, FGDC), the metadata version, and the field path from which each binding was drawn. The uncertainty obligations $U$ require a conflict flag when two standards supply incompatible values for the same field (e.g.\ inconsistent bounding boxes or access terms), and a coverage-completeness indicator when a required field is absent from all sources. The constraints $C$ restrict the spatial unit to county level
and restrict recency by a temporal threshold.

The preservation trace for a realistic realization would mark dataset identifiers and catalog source pointers as $\textsf{preserved}$, variable descriptions as $\textsf{approximated}$ when mapped across standards with differing vocabulary, access restrictions as $\textsf{weakened}$ when only a license name is available rather than a structured access-policy record, and any field absent from all consulted standards as $\textsf{unsupported}$, triggering a reported gap rather than a silent omission.

The contract structure is the same as for \textsc{ClaimConflict}: an answer-object kind, typed binding slots, required relations, witness obligations, provenance obligations, uncertainty obligations, and admissibility constraints. The regime shifts in exactly the same way. If the environment has no normalized catalog target, the query requires target construction. If the environment declares a \textsc{CatalogDiscovery} family with cross-standard field mappings, the query is partially specified and the system must instantiate and populate the target. This confirms that the formal vocabulary of Section~\ref{sec:target-contracts} is not specific to claim-bearing documents: it applies whenever a natural-language query requires an answer object whose obligations exceed the native result type of any single backend.

\section{Environment Classes and Regime Diagnosis}
\label{sec:environment-classes}

The previous sections defined target contracts and illustrated them through the \textsc{ClaimConflict} answer-object family. We now return to the broader data-management question: in what kinds of environments does target construction arise? The answer is not determined by the backend type alone. A relational database may participate in target construction if the query requires documents, extracted claims, and provenance. A document collection may support fixed-target querying if it already exposes a queryable claim graph or evidence language. A catalog query may be partially specified if the environment already declares a catalog-discovery target family, but target-construction if the relevant metadata commitments have not been normalized. An environment class is therefore not the same as a target-status regime. Regime membership is relative to the query and to the environment profile $\Gamma(E)$.

\subsection{Environment Class vs. Regime}

It is important not to associate fixed-target querying with structured databases and target construction with documents or heterogeneous systems. The same natural-language query can move across regimes as $\Gamma(E)$ changes. Consider again the query:
\begin{quote}
\textit{Find internal emails that contradict the claims of this announcement.}
\end{quote}
If the environment contains only raw documents, search indexes, and metadata, the system must construct a claim-conflict target before execution. If the environment already declares a \textsc{ClaimConflict} answer-object family and provides claim extraction, issue alignment, and contradiction scoring operators, the target family is known and the system must instantiate it. If the environment exposes a claim graph with queryable \texttt{contradicts} edges and span-level provenance, the same query may be fixed-target relative to that graph language.

The converse also holds. A query involving relational data may require target construction if the answer is not naturally a relation. A question asking whether a sequence of structured policy changes followed earlier concerns expressed in memos may require records, documents, event extraction, issue alignment, temporal ordering, and witnesses. Even if one component is relational, the target is a cross-source evidence object rather than a single SQL result. The regime is therefore determined by whether $\Gamma(E)$ already supports the obligations of an adequate target contract $\tau$.

\subsection{Environment Classes and Missing Target Commitments}

Table~\ref{tab:environment-regime-pressure} summarizes several common environment classes. The table does not classify environments by implementation technology alone. It asks what capabilities are usually available, what target commitments are often missing, and what kind of regime pressure the environment creates.

\begin{table*}[t]
\centering
\small
\begin{tabular}{|p{3.1cm}|p{4.3cm}|p{4.6cm}|p{4.1cm}|}
\hline
\textbf{Environment class} &
\textbf{Available capabilities} &
\textbf{Common missing target commitments} &
\textbf{Typical regime pressure} \\
\hline

Standard relational, graph, or RDF backend &
Schema or graph model, query language, execution engine, indexes, constraints &
Usually none for native lookup, join, aggregation, path, or triple-pattern tasks &
Fixed-target when the query's obligations are expressible in the declared language \\
\hline

Enriched single-model backend &
SQL, graph, or RDF language plus JSON, arrays, spatial functions, full-text predicates, vector procedures, or UDFs &
Dialect semantics, function grounding, operator signatures, portability across systems &
Usually fixed-target with extension grounding; target construction only when the answer object exceeds the backend model \\
\hline

Polystore or multi-model environment &
Multiple backends, wrappers, source mappings, cross-model access, possibly a federation or polystore engine &
Cross-source bindings, binding transfer, result reconciliation, answer assembly, provenance across models &
Partially specified or target-construction, depending on whether a composite target family is declared \\
\hline

Semantically related document collection &
Documents, metadata, search indexes, vector indexes, extraction layers, span anchors, entity links &
Claims, events, support/conflict relations, temporal relations, evidence witnesses, uncertainty, span provenance &
Partially specified when evidence-object families exist; target-construction when they must be introduced \\
\hline

Heterogeneous data catalog &
Dataset records, metadata fields, catalog schemas, metadata standards, search and ranking tools &
Normalized dataset, variable, spatial, temporal, license, access, quality, and lineage commitments across standards &
Partially specified when a catalog-discovery family exists; target-construction when the target vocabulary must be normalized \\
\hline

Workflow- or tool-mediated environment &
APIs, tools, extraction services, notebooks, analysis workflows, LLM tool calls, orchestration mechanisms &
Semantic target independent of action sequence, intermediate artifact obligations, failure modes, review conditions, provenance of tool outputs &
Partially specified or target-construction unless the target object is already declared independently of the workflow \\
\hline
\end{tabular}
\caption{Environment classes and the target-formation pressures they create. The regime is not determined by the environment class alone, but also by whether $\Gamma(E)$ supports the obligations of the target contract for the query.}
\label{tab:environment-regime-pressure}
\end{table*}

The first two rows explain why enriched backends do not automatically create a new target-construction problem. SQL with JSON, arrays, geospatial functions, or full-text predicates may be difficult to generate correctly, but it usually remains fixed-target if the desired answer is still a SQL-family result. Similarly, a graph query with spatial or vector procedures may require richer grounding, but the target family is still declared.

The later rows expose the shift addressed by \nlx. In a polystore, the question is often not just which backend language to generate, but what cross-source object is being requested. In a document collection, retrieval may be available while the evidentiary relation required by the user is not. In a heterogeneous catalog, search may be available while the normalized metadata commitments needed for the query are not. In a tool-mediated environment, a sequence of actions may be available while the semantic target that the actions are supposed to realize remains implicit.

\subsection{Regime Diagnosis by Environment Profile}

The fit test from Section~\ref{sec:target-contracts} provides a compact way to diagnose regimes. Given a candidate target contract $\tau$, each obligation $e \in \operatorname{Obl}(\tau)$ is checked against $\Gamma(E)$:
\[
\operatorname{fit}(e,\Gamma(E)) \in
\{
\textsf{exact},
\textsf{approximate},
\textsf{weakened},
\textsf{review},
\bot
\}.
\]
This does not produce a full natural-language understanding algorithm. It assumes that the system has proposed one or more candidate contracts and then asks whether the environment profile can support the obligations of those contracts. A useful diagnostic procedure is:

\begin{enumerate}
\item Identify the commitments induced by the query: bindings, relations, witnesses, provenance, uncertainty, and constraints.
\item Check whether a declared backend language in $\Gamma(E)$ can express and execute those commitments directly.
\item If not, check whether a declared answer-object family $K \in \mathcal{A}$ covers the commitments.
\item For each remaining obligation, compute or estimate its fit against the available sources, mappings, operators, and provenance mechanisms.
\item Classify the query as fixed-target, partially specified, or target-construction relative to the profile.
\end{enumerate}

A query is fixed-target only if the obligations of the adequate contract are captured by an existing backend or mediated language. It is partially specified if a declared answer-object family covers the obligations but the object must be instantiated through additional operations. It is target-construction if no declared backend language or answer-object family covers the obligation set.

\subsection{Examples Across Regimes}

Table~\ref{tab:regime-diagnosis-examples} illustrates how different queries and profiles induce different regimes. The examples are deliberately schematic; their purpose is to show how regime diagnosis follows from the relation between a query's obligations and $\Gamma(E)$.

\begin{table*}[t]
\centering
\small
\begin{tabular}{|p{4.4cm}|p{4.4cm}|p{3.0cm}|p{4.0cm}|}
\hline
\textbf{Query} &
\textbf{Environment profile} &
\textbf{Regime} &
\textbf{Reason} \\
\hline

\textit{List suppliers in California with contracts above \$1M.} &
Relational schema with supplier, contract, and location tables; SQL engine &
Fixed-target &
Bindings, predicates, joins, and projection are expressible in SQL. \\
\hline

\textit{Find emails that mention the product issue between 2019 and 2021.} &
Email archive with metadata fields, keyword index, and date filters &
Fixed-target &
The requested answer is a ranked or filtered document set supported by the retrieval interface. \\
\hline

\textit{Show how this product issue evolved across emails and reports.} &
Document collection with declared \textsc{IssueEvolutionTrace} family, issue mention extractor, timestamps, and source links &
Partially specified &
The answer-object family is known, but the trace must be instantiated and populated. \\
\hline

\textit{Find emails that contradict this announcement.} &
Raw announcement and email collections with search indexes but no claim-conflict family &
Target-construction &
The environment lacks a declared target for claims, issue alignment, contradiction, and paired witnesses. \\
\hline

\textit{Find emails that contradict this announcement.} &
Claim graph with queryable \texttt{contradicts} relation and span provenance &
Fixed-target &
The claim-conflict obligations are expressible through the declared graph language. \\
\hline

\textit{Find county-level wildfire smoke exposure datasets with variable descriptions and access restrictions.} &
Multiple catalogs with CKAN, schema.org, geospatial metadata, and local schemas, but no normalized catalog target &
Target-construction &
The system must first normalize dataset, variable, spatial, temporal, and access commitments across standards. \\
\hline

\textit{Find county-level wildfire smoke exposure datasets with variable descriptions and access restrictions.} &
Catalog environment with declared \textsc{CatalogDiscovery} family and cross-standard mappings &
Partially specified &
The target family is known; the system must instantiate it and rank datasets with provenance over metadata mappings. \\
\hline
\end{tabular}
\caption{Examples of regime diagnosis. The same natural-language query may fall into different regimes depending on the environment profile.}
\label{tab:regime-diagnosis-examples}
\end{table*}

The contradiction query appears twice in the table because it illustrates the central point. With only raw documents and retrieval, the target must be constructed. With a claim graph exposing contradiction as a queryable relation, the target is fixed. The string is the same; the regime changes because the profile changes.

The catalog query similarly changes regime. If metadata standards are unaligned, target construction is required to define the catalog commitments. If the environment already declares a catalog-discovery family with cross-standard mappings, the system can treat the query as partially specified and instantiate the existing target.

\subsection{Implications}

This profile-relative view has three implications. First, target construction is not a property of a data type: documents do not automatically imply target construction, and relational data do not automatically imply fixed-target translation. The relevant question is whether the profile supports the obligations of the target contract. Second, target-definition infrastructure should evolve: when repeated queries require new bindings, relations, witnesses, or provenance conventions, the system should update $\Gamma(E)$ by adding mappings, operators, answer-object families, or provenance mechanisms, so that a query requiring target construction early in the life of a system may later become partially specified or fixed-target. Third, evaluation should record the profile state under which a query was interpreted: a system's target choice is not reproducible unless the benchmark or execution record includes the relevant environment profile, and the same answer may be reasonable under one profile and inadequate under another.

The next section turns from diagnosis to research directions. It asks what system components would be needed to make environment profiles, target contracts, realization plans, preservation traces, and target-fidelity benchmarks reusable artifacts rather than ad hoc products of individual prompts or workflows.

\section{Research Agenda: Target Definition as a Systems Layer}
\label{sec:terrain}

The preceding sections argued that target formation should be treated as a first-class systems problem. The central object is a target-definition layer that sits between natural-language interpretation and backend execution. In simple settings, this layer may collapse into a schema and a known query language. In heterogeneous settings, it must expose what the environment can support, what answer target is adequate, how that target can be realized, and what obligations are preserved or degraded during realization. The formal spine developed earlier suggests a corresponding systems agenda:
\[
\Gamma(E)
\rightsquigarrow \tau
\rightsquigarrow \pi
\rightsquigarrow (\rho, a).
\]
Environment profiles $\Gamma(E)$ support target diagnosis. Target contracts $\tau$ specify answer objects and their obligations. Realization plans $\pi$ compile target contracts into executable operations. Preservation traces $\rho$ record whether obligations survive realization. Answer objects $a$ provide the returned structure that users and evaluators inspect. The research directions below ask how these artifacts can become reusable, maintainable, and benchmarkable system components.

\begin{table*}[t]
\centering
\small
\begin{tabular}{|p{2.8cm}|p{4.0cm}|p{4.4cm}|p{4.4cm}|}
\hline
\textbf{Formal object} &
\textbf{Systems problem} &
\textbf{Needed artifact} &
\textbf{Evaluation question} \\
\hline
$\Gamma(E)$ &
Environment profiling and capability diagnosis &
Versioned profile of sources, mappings, operators, answer families, provenance and uncertainty mechanisms &
Can the system correctly diagnose whether a query is fixed-target, partially specified, or target-construction? \\
\hline
$\tau$ &
Target-contract derivation &
Inspectable contract specifying answer kind, bindings, relations, witnesses, provenance, uncertainty, and constraints &
Does the contract capture the user's information need and the obligations needed for evaluation? \\
\hline
$K \in \mathcal{A}$ &
Governed answer-object libraries &
Reusable families such as conflict, comparison, timeline, evidence bundle, catalog-discovery target, and workflow-backed report &
Can a small library cover recurring workloads without degenerating into ad hoc templates? \\
\hline
$\pi$ &
Realization planning &
Operator plan assigning target obligations to retrieval, extraction, query, mapping, ranking, review, or assembly steps &
Does the plan cover the contract obligations using available capabilities? \\
\hline
$\rho$ &
Preservation tracing &
Trace over target obligations indicating preserved, approximated, weakened, or review-dependent status &
Does the trace correctly expose degradation rather than hiding it behind the final answer? \\
\hline
$a \models \tau$ &
Target-fidelity evaluation &
Benchmark instances with queries, profiles, contracts, expected witnesses, provenance requirements, and acceptable answer objects &
Can evaluation distinguish fluent answers from contract-satisfying or trace-qualified answers? \\
\hline
\end{tabular}
\caption{Research agenda induced by the target-definition layer. Each direction corresponds to one artifact in the target-aware querying spine.}
\label{tab:research-agenda-spine}
\end{table*}

\subsection{Direction 1: Environment Profiles and Capability Diagnosis}

The first research direction is to design environment profiles that support target diagnosis. A profile should not merely list sources. It should describe the target-relevant capabilities of an environment: available data sources, source descriptions, mappings, operators, answer-object families, provenance mechanisms, uncertainty mechanisms, and known realization pathways. This differs from ordinary cataloging. A conventional catalog may describe datasets, schemas, access locations, or metadata fields. A target-aware profile must also say what kinds of answer objects the environment can support. For a polystore, this includes cross-model mappings, binding-transfer capabilities, and provenance conventions. For a document collection, it includes span anchors, extraction products, claim or event layers, and evidence relations. For a metadata catalog, it includes cross-standard mappings for dataset identity, variables, spatial coverage, temporal coverage, licensing, access restrictions, lineage, and quality indicators.

\noindent\textbf{Concrete problem.}
Given a query $q$, an environment profile $\Gamma(E)$, and one or more candidate target contracts, determine whether the environment supports fixed-target translation, partially specified target completion, or target construction. When support fails, identify the missing capability: a source, mapping, operator, answer-object family, provenance mechanism, uncertainty mechanism, or realization pathway.

\noindent\textbf{Nature of a solution.}
A solution would provide a versioned profile model and a diagnostic procedure based on contract fit. For each obligation $e \in \operatorname{Obl}(\tau)$, the system would estimate whether the environment supports it exactly, approximately, weakly, through review, or not at all. Failed checks would generate profile-repair tasks, such as registering a new extractor, adding a mapping, declaring a new answer-object family, exposing a provenance source, or requesting human curation.

\noindent\textbf{Success indicators.}
Success would be measured by diagnostic accuracy and profile usefulness. A system should correctly distinguish fixed-target, partially specified, and target-construction cases. It should identify missing capabilities when a target cannot be realized. It should also support reproducibility: the system should be able to record which version of $\Gamma(E)$ was used to interpret a query and why a later profile might change the regime assignment.

\subsection{Direction 2: Target-Contract Derivation}

The second research direction is to derive target contracts from natural-language questions. The formalism treats $\tau$ as the object that specifies what an adequate answer must contain. A practical system must infer or construct such a contract from a query, the interaction context, the environment profile, and possibly prior workload examples. The system must decide what type of answer object is being requested. A query asking for ``emails that contradict this announcement'' may require a claim-conflict contract. A query asking ``how did this issue evolve?'' may require an issue-evolution trace. A query asking for datasets with spatial coverage, variables, and access restrictions may require a catalog-discovery contract. In each case, the output should be an inspectable target contract rather than an opaque prompt or tool sequence.

\noindent\textbf{Concrete problem.}
Given a natural-language query $q$ and an environment profile $\Gamma(E)$, infer an adequate target contract
\begin{quote}\centering
$\tau = \langle K, B, R, W, P, U, C \rangle$.
\end{quote}
The system must identify the answer-object kind, required bindings, relations, witness obligations, provenance obligations, uncertainty requirements, and admissibility constraints. It must also detect ambiguity: multiple contracts may be plausible, or the query may require clarification before a contract can be selected.

\noindent\textbf{Nature of a solution.}
A solution may combine semantic parsing, LLM-based proposal generation, workload templates, interaction policies, and profile checking. One architecture is to generate candidate contracts, test them against $\Gamma(E)$, and either select the best-supported contract, revise it, or ask a focused clarification question. Expert-authored answer-object families and examples can constrain LLM-generated contracts so that the output remains inspectable and reusable.

\noindent\textbf{Success indicators.}
Success would be measured by contract adequacy and downstream usefulness. Expert annotators should agree that the derived contract captures the user's information need. The contract should expose the obligations needed for evaluation. Errors in later realization should be traceable to specific contract obligations or missing environmental capabilities rather than appearing only as opaque answer failures.

\subsection{Direction 3: Governed Answer-Object Libraries}

The third research direction is to build governed libraries of answer-object families. Fixed-target systems often inherit their answer form from the backend language: tuples, triples, graph paths, ranked documents, or aggregate values. Target-aware systems need reusable answer-object families that are not native to any one backend. Examples include claim-conflict objects, evidence bundles, comparison objects, issue-evolution traces, timelines, catalog-discovery targets, and workflow-backed reports.

The challenge is governance. If every query produces a new hand-crafted template, target contracts become prompt artifacts rather than system artifacts. A useful library should contain a small set of reusable families, together with specialization and composition rules. A claim-conflict object may specialize a comparison family. An issue-evolution trace may compose a timeline with an evidence bundle. A catalog-discovery target may specialize into a geospatial-data discovery target when spatial coverage becomes a required commitment.

\noindent\textbf{Concrete problem.}
Determine whether a compact library of answer-object families can cover a substantial fraction of recurring workloads in heterogeneous data environments. Each family should specify its contract schema: slots, relation types, witness requirements, provenance requirements, uncertainty annotations, admissibility constraints, and expected realization strategies.

\noindent\textbf{Nature of a solution.}
A solution would treat answer-object families as versioned, parameterized system artifacts. Families would have declared slots and constraints, but also extension points. The library would support specialization, composition, and reuse. Workload analysis could identify recurring families, while expert curation would prevent uncontrolled proliferation. The library should also record examples and counterexamples so that systems can learn when a query fits an existing family and when a new family is justified.

\noindent\textbf{Success indicators.}
Success would be measured by coverage, reuse, and consistency. A small number of families should cover many realistic target-construction queries. Experts should agree that queries assigned to a family genuinely instantiate that family. Systems using the library should produce more stable and inspectable answer structures than systems that construct answer formats ad hoc.

\subsection{Direction 4: Realization Planning and Preservation Traces}

The fourth research direction is to compile target contracts into executable realization plans while preserving contract obligations. Once the system has selected a target contract $\tau$, it must construct a plan $\pi$ over the capabilities in $\Gamma(E)$. In heterogeneous settings, this plan may combine retrieval, extraction, SQL, graph traversal, metadata mapping, entity resolution, ranking, model calls, answer assembly, and review.

This planning problem differs from ordinary query optimization. The object being planned is not only an algebraic query, but a target contract with obligations. A retrieval operator may find candidate documents but not discharge a witness obligation. An extraction operator may introduce claim spans but only approximately. A metadata lookup may preserve timestamps exactly. A review step may discharge a relation that cannot be safely asserted automatically. The planner must therefore reason about cost, capability, and preservation together.

\noindent\textbf{Concrete problem.}
Given a target contract $\tau$ and an environment profile $\Gamma(E)$, construct a realization plan $\pi$ that covers the obligations in $\operatorname{Obl}(\tau)$ and produces a preservation trace $\rho$. The system must decide which obligations are handled by which operators, which obligations are exact, which are approximate, which are weakened, and which require review.

\noindent\textbf{Nature of a solution.}
A solution would extend query and workflow planning with contract-aware operator descriptions. Operators would advertise not only input-output types and costs, but also their effects on target obligations. For example, a span extractor may introduce candidate spans and preserve offsets when the source format supports them. A vector retrieval operator may support candidate generation but only approximate relevance. A contradiction classifier may support a relation obligation but require review below a confidence threshold. Plans would carry preservation traces as first-class outputs.

\noindent\textbf{Success indicators.}
Success would be measured by realization quality and trace fidelity. Plans should cover more contract obligations, preserve more witnesses and provenance, and expose more degradations than ad hoc tool orchestration. Preservation traces should help users and developers locate answer failures --- missing witnesses, weak entity alignment, unsupported conflict claims, or dropped provenance. A strong system should avoid certifying an answer as satisfying a target contract when important obligations were only approximated, weakened, or left for review.

\subsection{Direction 5: Target-Fidelity Benchmarks}

The final research direction is evaluation. Existing NL-to-SQL benchmarks evaluate generated queries or execution results. Retrieval benchmarks evaluate ranked lists. RAG benchmarks often evaluate answer quality, citation quality, or faithfulness. These are important, but they do not evaluate whether the system constructed the right target or preserved its obligations.

A target-fidelity benchmark should include the intermediate artifacts needed to judge target formation. A benchmark instance should include a query, an environment profile, available answer-object families, one or more acceptable target contracts, expected witness structures, provenance requirements, uncertainty expectations, and expected preservation behavior. The final answer can then be evaluated together with the target contract and trace.

\noindent\textbf{Concrete problem.}
Define benchmark instances of the form
\[
\langle q, \Gamma(E), \mathcal{A}, \mathcal{T}^{\mathsf{acc}},
\mathcal{W}^{\mathsf{acc}}, \mathcal{P}^{\mathsf{acc}},
\mathcal{R}^{\mathsf{acc}} \rangle,
\]
where $q$ is the natural-language query, $\Gamma(E)$ is the environment profile, $\mathcal{A}$ is the available answer-object library, $\mathcal{T}^{\mathsf{acc}}$ is the set of acceptable target contracts, $\mathcal{W}^{\mathsf{acc}}$ and $\mathcal{P}^{\mathsf{acc}}$ specify expected witness and provenance obligations, and $\mathcal{R}^{\mathsf{acc}}$ specifies expected preservation or review behaviour.

\noindent\textbf{Nature of a solution.}
A solution would decompose evaluation into subtasks corresponding to
the formal spine.
(i)\textit{Contract adequacy}: does the derived contract capture the bindings, relations, witnesses, provenance, and uncertainty that expert annotators associate with the query?
(ii)\textit{Trace fidelity}: does $\rho$ correctly record $\textsf{preserved}$, $\textsf{approximated}$, $\textsf{weakened}$, and $\textsf{requiresReview}$ statuses, or does it silently upgrade degraded obligations?
(iii)~\textit{Answer-object correctness}: does $a$ fill the required slots, represent the required relations, and attach required witnesses and provenance?

Because more than one contract may be adequate for a query, the benchmark should define equivalence classes of contracts rather than a single gold standard. A preliminary version of such a benchmark is tractable with a small workload: ten to twenty queries spanning fixed-target, partially specified, and target-construction cases, each annotated with an environment profile, an acceptable contract set, expected witnesses,
expected provenance obligations, and expected trace statuses. Even a benchmark of this size would already distinguish systems that construct inspectable targets from systems that produce fluent answers without target formation. Scaling to larger workloads requires semi-automatic contract annotation, which is a hard open problem but one that is no harder than schema annotation in existing NL2SQL benchmarks.

\noindent\textbf{Success indicators.}
A target-fidelity benchmark succeeds if its subtask scores predict expert judgments of trustworthiness and auditability better than surface-level fluency scores do. Concretely: a system that correctly identifies the target-construction regime, derives an adequate contract, and returns a trace-qualified answer with explicit degradation records should score higher than a system that returns a fluent, relevant, but structurally unaccountable answer --- even if a naive human reader initially prefers the latter. Progress on subtasks (i) through (iii) should also be independently
measurable, so that benchmark improvements can be localized to specific components of the target-definition layer rather than attributed only to overall answer quality.

\subsection{Summary}

The research directions above define a target-definition layer for natural-language data systems. Environment profiles describe what can be supported. Target-contract derivation specifies what should be produced. Answer-object libraries make recurring targets reusable. Realization planning compiles contracts into executable operations. Preservation traces expose what was preserved or degraded. Target-fidelity benchmarks make the process evaluable.

The long-term goal is a class of target-aware data systems that would not merely translate natural language into a query, retrieve relevant passages, or call tools. They would expose the target they selected or constructed, explain why it is appropriate for the query and environment, realize it through available capabilities, and report whether its obligations were preserved, approximated, weakened, or left for review.

\section{Related Research}
\label{sec:related}

The \nlx viewpoint proposed in this vision paper builds on several long-standing threads in database and AI research: natural-language interfaces, keyword search, data integration, polystores, document extraction, provenance, semantic operators, and tool-using language-model systems. The contribution is that we isolate the \textit{target definition} tasks: the question of whether a natural-language query should be translated into a fixed backend language, completed within a declared answer-object family, or used to form a new target contract before backend-specific execution
begins.

\subsection{Fixed-Target Natural-Language Interfaces}

Natural-language interfaces to databases have been studied for decades, and much of the modern literature fits the fixed-target regime. In relational settings, NLIDB and text-to-SQL systems assume that SQL is the destination language: the central problem is to
ground user language in the database schema, infer joins and predicates, and synthesize an executable query \cite{ozcan2020nli2data,kim2020natural,li2014constructing}.
Comparable assumptions appear in NL interfaces to graph and RDF systems, where the target language is Cypher, GQL, or SPARQL \cite{tsampos2025domain,liang2024aligning}. Such systems are not displaced by \nlx. They occupy the regime in which target identification has already been resolved. The \nlx question deals with what happens when the backend language, answer form, or target object is already inadequate for the user query.

\subsection{Keyword Search and Schema-Agnostic Access}

Keyword search over structured data is an important precursor to
\nlx because it separates the user's access language from the
underlying query language and schema. Systems such as
\textsc{DBXplorer}, \textsc{DISCOVER}, and \textsc{BANKS} allow
users to access relational data without writing SQL, using keyword
matching, join discovery, graph traversal, and ranked connected
answers \cite{agrawal2002dbxplorer,hristidis2002discover,aditya2002banks}.
These systems are closer to \nlx than ordinary fixed-target
translation because they construct intermediate result structures on
the user's behalf rather than simply translating user input into a
declared query. However, the family of structures to be constructed
is still largely fixed by the system---tuple trees, candidate
joining networks, or ranked connected answers---and the system does
not ask whether the user's information need requires a different
answer-object family such as a claim-conflict object, evidence
bundle, or cross-source witness.

\subsection{Data Integration, Polystores, Heterogeneous Querying,
and Catalog Discovery}

The database lineage closest to \nlx is data integration over
heterogeneous sources. Mediator-wrapper systems and foundational
work on answering queries using views studied access to multiple
sources through a mediated schema, wrappers, mappings, and query
decomposition
\cite{hammer1995tsimmis,carey1995garlic,levy1995views,lenzerini2002data}.
Dataspace systems later weakened the requirement of complete upfront
integration and emphasized pay-as-you-go coordination over
heterogeneous sources \cite{franklin2005dataspaces,halevy2006principles}.
Polystore and cross-platform systems extend this tradition by
coordinating multiple execution engines, assigning subtasks to
appropriate engines and planning data movement between them
\cite{kruse2018rheemix}. These systems make the target-definition
problem more visible, because a user request may require several
sources and operators before an answer can be produced.

Recent work has also begun to study natural-language querying beyond
a single relational backend. Work on declarative techniques for
heterogeneous NL queries studies questions whose answers require
coordinated database and API calls rather than a single SQL query
\cite{khabiri2025declarative}. Multi-model benchmarks such as SM3
evaluate NL questions over relational, document, graph, and RDF
representations with target queries expressed in SQL, MQL, Cypher,
and SPARQL \cite{sivasubramaniam2024sm3}. Work on multi-database
Text2Cypher and enterprise database routing treats source selection
and decomposition as part of NL data access
\cite{ozsoy2026multidatabase,sudarshan2026routing}. Meta Engine
coordinates heterogeneous LLM-based semantic query systems using an
NL query parser, operator generator, query router, adapters, and
result aggregator \cite{li2026metaengine}. This recent work moves
beyond the simplest fixed-target setting and is closer to \nlx than
ordinary NL2SQL.

However, most of this work still assumes that the target form is already given at some level: a mediated schema, ontology, API or database call set, one of several known backend query languages, or a query-system integration substrate. This observation applies in particular to ontology-mediated query answering (OMQA) and ontology-based data access (OBDA), which are the traditions most structurally similar to \nlx. In OMQA/OBDA, a domain ontology mediates between user queries, expressed in a description logic or similar formalism, and heterogeneous sources accessed through mappings. The ontology plays a normalization role: it declares the concepts, roles, and individuals that queries range over, and the mapping layer connects that vocabulary to source schemas. This is a powerful architecture for queries whose answer type is already covered by the ontological vocabulary. The \nlx distinction arises precisely when that coverage fails. A natural-language query asking for internal emails that contradict a public announcement cannot be answered by selecting from a domain ontology that declares entities, properties, and class memberships. It requires an answer-object family that binds claims, conflict relations, source-span witnesses, and uncertainty --- none of which are obligatory components of standard OBDA ontologies or mapping languages. A target contract in \nlx is not an ontological concept definition or a GAV/LAV mapping rule. It is a specification of what a returned answer object must contain and what obligations realization must preserve. The OBDA machinery --- ontologies, mappings, query rewriting, and source access --- may contribute operators during realization, but the target contract is defined upstream of and independently of any particular ontological vocabulary.

Data lake and catalog systems address a related problem: how users
find, relate, and reuse datasets in large heterogeneous repositories.
Systems such as \textsc{Aurum} and recent work on unified data
discovery support dataset search, relationship discovery, and
heterogeneous query modalities across structured, semi-structured,
and multimodal sources
\cite{fernandez2018aurum,wang2026unidisc,wang2025heterogeneous}.
These systems provide important machinery for finding and navigating
large collections, but they typically assume the family of returned
objects: datasets, tables, columns, or ranked catalog records. \nlx
asks an earlier target-definition question. For a natural-language
request such as \textit{``Find recent county-level datasets on
wildfire smoke exposure with spatial coverage, variable descriptions,
and access restrictions,''} the system must determine which metadata
commitments define an adequate answer---binding datasets, variables,
spatial coverage, temporal extent, access policy, lineage, and
supporting metadata fields across CKAN, schema.org, DCAT, FGDC, and
related standards. \nlx treats the catalog answer itself as a target
object and asks whether it preserves the metadata commitments
required by the natural-language question.

\subsection{Document Extraction, Evidence, and Provenance}

Document-centric database research provides an important foundation
for \nlx because many target-construction queries require
evidence-bearing structures rather than only retrieved documents.
Declarative information extraction systems such as \textsc{SystemT}
showed that weakly structured text can be queried by defining
extraction layers over spans and derived relations
\cite{krishnamurthy2008systemt,chiticariu2010systemt}. The theory of
document spanners later gave a formal account of extraction as
mappings from documents to relations over spans, with relational
operators over extracted span relations
\cite{fagin2015relational,peterfreund2018recursive}. This line of
work shows that document querying often requires an intermediate
structured layer before database-style querying becomes meaningful.

Recent systems revisit this problem in the setting of AI-assisted
extraction. \textsc{DocETL} studies declarative pipelines for
complex document processing and uses agentic rewrites and evaluation
mechanisms to improve LLM-powered extraction quality
\cite{shankar2024docetl}. Domain-specific systems for auditable
evidence extraction further show the importance of typed schemas,
evidence-gated decisions, conflict-aware consolidation, and
sentence-level provenance in high-stakes document analysis
\cite{mortezaagha2026chaos}. Provenance research is central to this
discussion because many \nlx targets require witnesses, not merely
answers. Classical database work studied why-, where-, and
how-provenance and algebraic models for propagating annotations
through queries \cite{buneman2001why,green2007provenance}, and later
work developed provenance as a basis for explanation, debugging,
trust, and reproducibility \cite{cheney2009provenance,bourhis2020equivalence}.

Extraction systems, spanners, RAG pipelines, and provenance
frameworks usually assume that the extraction schema, query,
workflow, or answer format has already been selected. \nlx asks an
earlier target-definition question: what answer object should the
natural-language query require in the first place? For a
claim-conflict query, the target contract requires reference claims,
candidate claims, issue alignment, contradiction relations, paired
source-span witnesses, provenance, uncertainty, and review
status---not simply a retrieval result or a generated answer with
citations. Document extraction and provenance are not replaced by
\nlx; they are components used during realization. The \nlx
contribution is to move witness and provenance requirements into the
target contract itself, so that the preservation trace records
whether those obligations survived realization, were approximated,
were weakened, or remained unresolved.

\subsection{Semantic Operators, RAG, and Tool-Mediated Semantic Plans}

Recent systems work has moved toward treating language-model-based
operations as first-class data-processing constructs.
Semantic-operator systems extend database or DataFrame-style
interfaces with AI-powered predicates, joins, mappings, rankings,
aggregations, and classifiers. \textsc{LOTUS} introduces semantic
operators for analytics over text data, with multiple physical
implementations and optimizations for semantic filters, clustering,
and joins \cite{patel2024semanticops}. \textsc{Palimpzest} treats
AI-powered analytics over unstructured data as a declarative
optimization problem, choosing plans that trade off quality, runtime,
and cost \cite{liu2024palimpzest}. \textsc{Sema} embeds
natural-language expressions inside SQL clauses and optimizes under
latency, cost, and accuracy constraints \cite{qi2026sema}. Semantic
integrity constraints provide declarative guardrails for
AI-augmented data processing, including grounding, soundness, and
exclusion constraints over LLM outputs \cite{lee2025sic}. These
systems are close to \nlx because they expose semantic operations
and query-processing structure rather than treating the LLM as a
black box.

Retrieval-augmented generation systems retrieve external evidence before generation; recent work has argued that RAG architectures can be strengthened through views, query analysis, planning, and
provenance \cite{tan2023reimagining}, while broader surveys describe
increasingly modular retrieval, augmentation, and generation
pipelines \cite{gao2023rag}. Tool-using language-model systems such
as \textsc{ReAct} and \textsc{Toolformer} make action selection and
external tool invocation part of language-model inference
\cite{yao2022react,schick2023toolformer}. These lines of work are
among the closest neighbors to \nlx because they make intermediate
operations explicit: semantic filters, semantic joins, retrieval
plans, tool calls, query rewrites, routing decisions, and
cost-quality choices. However, they typically assume that the
relevant target form has already been chosen: a semantic-operator
program, a SQL extension, a RAG plan, a tool-action sequence, or an
API-mediated workflow.

The \nlx distinction is to place target formation before those execution choices. A natural-language query may be realizable through semantic operators, RAG, or tool calls, but the system must first determine what answer object those operations are meant to produce. For a claim-conflict query, the target is a claim-conflict contract with bindings, relations, witnesses, provenance, uncertainty, and certification conditions. The same target may be realized by several semantic plans, and the same plan may be inadequate if it fails to preserve the target's obligations. Semantic operators and tool-mediated plans are realization mechanisms; \nlx asks which target contract they should realize.

\begin{table*}[t]
\centering
\small
\begin{tabular}{|p{3.2cm}|p{4.8cm}|p{7.0cm}|}
\hline
\textbf{Existing abstraction} &
\textbf{What it typically assumes} &
\textbf{What \nlx adds} \\
\hline
NL2SQL, NL2GQL, NL2SPARQL &
A backend query language is fixed in advance. &
Target-status diagnosis: the system must decide whether fixed-target translation is adequate. \\
\hline
Keyword search over structured data &
The answer is a tuple, tuple tree, connected structure, or ranked result. &
Answer-object formation when the user needs evidence, comparison, conflict, or cross-source witnesses. \\
\hline
Mediated schema, data integration, OBDA &
A mediated vocabulary, ontology, or mapping layer exists. &
Target vocabulary or answer-object family may itself need to be selected or constructed. \\
\hline
Polystore and cross-platform optimization &
A logical task is available for decomposition across engines. &
A target contract must precede decomposition and specify what obligations execution must preserve. \\
\hline
Data lake and catalog discovery &
Datasets and metadata can be searched, organized, or related. &
Natural-language catalog discovery may require normalized target contracts over cross-standard metadata commitments. \\
\hline
Document extraction and spanners &
An extraction schema or span relation is specified. &
The answer-object family may need to be selected or constructed before extraction is meaningful. \\
\hline
Provenance and lineage &
A query, workflow, or computation has been specified. &
Witness and provenance obligations become part of the target contract, not only post-hoc annotations. \\
\hline
Semantic operators and AI-augmented query systems &
A semantic operator language, API, or workflow model is available. &
Target contracts precede operator choice and specify what semantic obligations operators must preserve. \\
\hline
RAG and tool-using agents &
Retrieval, tools, or action plans may be selected implicitly. &
The target and its preservation trace are explicit, inspectable, and evaluable. \\
\hline
\end{tabular}
\caption{Positioning \nlx relative to related research traditions. The gap addressed by this paper is target definition: deciding whether a natural-language query should be translated into a fixed backend language, completed within a declared answer-object family, or used to construct a new target contract before execution.}
\label{tab:related-positioning}
\end{table*}

\subsection{\nlx in Perspective}

Table~\ref{tab:related-positioning} summarizes the relationship between \nlx and the main research traditions discussed above. Most of these traditions begin after some target vocabulary, schema, operator language, or tool space has already been selected, and together they provide many components needed for target-aware natural-language data systems: translation, mediation, rewriting, extraction, discovery, semantic operators, provenance tracking, and tool orchestration. The missing layer is a target-definition
framework that makes explicit what answer object is to be formed, why it is adequate for the query and environment, how it is realized, and what obligations must be preserved.

\section{Conclusion}
\label{sec:conclusion}

Natural-language interfaces to data have largely been built around a powerful assumption: the target of interpretation is already known. That assumption remains appropriate for many NL2SQL, NL-to-GQL, NL-to-SPARQL, retrieval, and tool-calling settings. But it is insufficient for heterogeneous environments in which users may be asking not for a tuple set, graph path, ranked document list, or API result, but for an answer object: an evidence structure, comparison, timeline, conflict, catalog-discovery object, or cross-source witness.

This paper has proposed the \nlx lens to make that prior question explicit. The central issue is whether the target itself is fixed, partially specified, or must be constructed. Fixed-target querying remains an important limiting case. Partially specified regimes arise when the answer-object family is known but the particular object must still be instantiated. Target-construction regimes arise when the environment does not yet provide an adequate target family, evidence structure, or normalized vocabulary.

The paper developed a compact systems vocabulary for this problem via the concepts of environment profile, target contract, realization plan, and preservation trace. The returned answer object can then be evaluated not only for plausibility, but for whether it satisfies or trace-qualifies the target contract. The claim-conflict example illustrates why this decomposition matters. A query asking for internal emails that contradict an announcement is not adequately answered by retrieval alone. It requires reference claims, candidate claims, issue alignment, a conflict relation, source-span witnesses, provenance, uncertainty, and possibly review. The target contract makes those obligations explicit before the system chooses whether to use retrieval, extraction, graph queries, model scoring, metadata lookup, or human validation.

The broader implication is that future natural-language data systems need a target-definition layer. Such a layer would maintain environment profiles, derive target contracts, govern reusable answer-object families, plan realizations over heterogeneous capabilities, and record preservation traces. It would also change evaluation: benchmarks should ask not only whether a final answer is fluent or executable, but whether the system selected the right target status, constructed an adequate target, preserved witnesses and provenance, exposed uncertainty, and avoided unreported degradation.

\bibliographystyle{ACM-Reference-Format}
\bibliography{refs}

\end{document}